\documentclass[letterpaper, 10 pt, conference]{ieeeconf}  
\usepackage{amsmath}
\usepackage{amssymb}
\usepackage{graphicx}
\usepackage{cite}
\usepackage{bm}
\usepackage{caption}
\usepackage{comment}
\usepackage{color}
\usepackage{algorithmicx}
\usepackage{amsfonts}
\usepackage{subcaption}
\usepackage[ruled,vlined,linesnumbered,noresetcount]{algorithm2e}
\usepackage{array}
\usepackage[font=small]{caption}

\usepackage[export]{adjustbox}

\IEEEoverridecommandlockouts 
\title{\LARGE \bf
MPC-based Vibration Control and Energy Harvesting Using an Electromagnetic Vibration Absorber With Inertia Nonlinearity
}

\author{Kaian Chen, Zhaojian Li, Wei-Che Tai, Kai Wu and Yan Wang
\thanks{Kaian Chen, Zhaojian Li, and Wei-Che Tai are with the department of Mechanical Engineering, Michigan State University, East Lansing, MI, 48824, USA. Email:
        {\tt\small \{chenkaia,lizhaoj1,taiweich\}@msu.edu}}%
\thanks{Kai Wu and Yan Wang are with Research \& Advanced Engineering,
 Ford Motor Company, Dearborn, MI 48121, USA.
        Email: {\tt\small \{kwu41,ywang21\}@ford.com}}
}

\begin{document}
\maketitle
\thispagestyle{empty}
\pagestyle{empty}

\begin{abstract}
Simultaneous vibration control and energy harvesting of vehicle suspensions has attracted great research interests over the past decades. However, existing frameworks tradeoff suspension performance for energy recovery and are only responsive to  narrow-bandwidth vibrations.  In this paper, a new energy-regenerative vibration absorber (ERVA) using a ball-screw mechanism is investigated. The ERVA system is based on a rotary electromagnetic generator with adjustable nonlinear rotational inertia which  passively increases the moment of inertia as the vibration amplitude increases. This structure is effective for energy harvesting and vibration control without increasing the suspension size. Furthermore, a nonlinear model predictive controller (NMPC) is applied to the system for further performance enhancement  where we  exploit  road profile information as a preview. The performance of NMPC-based ERVA is demonstrated in a number of simulations and superior performance is demonstrated.
\end{abstract}

\section{Introduction}\label{sec:intro}

Semi-active suspensions with adjustable damping were introduced in 1970s and can provide better trade-offs between suspension performance and power consumption than passive or active suspensions \cite{deakin2000,dogruer2008}. For example, studies show that magnetorheological (MR) controllable fluid dampers demonstrated significantly improved ride quality and road handling with only hundreds of watts of actuation power \cite{wray2003,wray2005}. However, the need of electricity tethers active and semi-active suspensions to external batteries, increasing design difficulties and decreasing system reliability. Meanwhile, there is a fair amount of energy being dissipated by vehicle shock absorbers. Depending on the vehicle type, it is estimated that 100 watts to 10 kilo-watts of electrical power can be harvested from vehicle suspensions \cite{abdelkareem2018}. These energy harvesting potentials offer an invaluable energy source to power the controller.

As such, energy-regenerative suspensions that can convert vibration energy into electricity to self-power the controllers have been widely studied in the past decades. Different transducers have been used to convert the energy, including magnetorheological transducers \cite{choi2009,chen2012,sapinski2014}, hydraulic-electromagnetic transducers \cite{fang2013,xu2015}, and linear \cite{goldner2001,gupta2006,zhang2007,zuo2010,tang2014} and rotary electromagnetic transducers \cite{li2012,li2013,maravandi2015}. In this work, we will focus on electromagnetic transducers. In particular, the linear electromagnetic transducers directly convert the linear reciprocal vibration motion of the suspension into electricity, using a set of voice coils moving in an array of magnetic fields. The damping force is then generated due to the back electromotive force \cite{karnopp1989,martins2006}. Yet, they typically have small energy density and bulky designs. For example, the linear regenerative shock absorber patented by Tufts University \cite{goldner2001} weights over 150 pounds and has a too large diameter (6") to fit into most existing vehicle suspensions. The rotary systems, on the other hand, utilize certain linear-rotational motion transmission mechanisms, such as a ball screw or rack-pinion, to transmit the reciprocating linear vibration into rotational motion, driving a rotary electromagnetic transducer for electricity. Because the motion transmission ratio can be amplified, e.g., by gearboxes, the energy density and damping force can be much larger than the linear systems. Furthermore, one can achieve semi-active damping control by varying the duty cycle of an electromagnetic transducer \cite{okada2002variable} or active damping control by applying a drive voltage to the transducer to generate control input \cite{kawamoto2008}.

The rotary systems can be readily integrated with the inerter, which is a new passive vibration control device recently introduced to achieve better suspension performance for vehicles \cite{smith2002,smith2004,papageorgiou2005,papageorgiou2006,wang2008,smith2009,wang2011,hu2014}. Similar to the rotary systems, the inerter typically uses a rack-pinion \cite{wang2009} or ball screw mechanism \cite{ikago2012} to convert the linear suspension vibration to rotational motion of an object with large moment of inertia, e.g., a flywheel. Due to the large moment of inertia, an inerter device with small physical mass can provide a large inertia force equivalent to the inertia force of a translating object with ten to hundreds times physical mass \cite{ikago2012,papageorgiou2005}. By attaching a flywheel, the rotary systems then can have dual capabilities of energy regeneration and semi-active vibration control. 

Motivated by these advantages, a new energy-regenerative vibration absorber (ERVA) with nonlinear inertia is investigated. The absorber consists of a rotary electromagnetic generator and a nonlinear rotational inertia vibration absorber that passively increases its moment of inertia as the vibration amplitude increases. By doing so, the effectiveness of energy harvesting and vibration control are simultaneously enhanced when compared with conventional linear inerters that have a constant moment of inertia. Furthermore, the damping force of the absorber is adjustable to achieve semi-active damping control.

To further improve the energy harvesting efficiency while maintaining good ride comfort, a nonlinear model predictive controller (NMPC) is developed to adjust energy harvesting circuit by optimizing the process according to external information, i.e. the road profile. With the advent of technologies that  use existing vehicle sensors for road profile estimation \cite{profile1,profile2}, the proposed preview control with NMPC is practically appealing.  This combined strategy, NMPC embedded in the novel ERVA, can provide adjustable vibration control and energy harvesting according to specific requirements, minimization of chassis acceleration for driving comfort, maximization of energy harvesting, or a trade-off for both satisfaction.

The rest of this paper is organized as follows. In section \ref{sec:QCar_Model}, the structure of ERVA is introduced. Section \ref{sec:Formulation} describes the system formulation, including a simplified linear benchmark model, as well as the NMPC implementation. A set of simulation results and performance evaluation are presented in section \ref{sec:Sim}. The section \ref{sec:Conclusion} concludes the paper.

\section{Vibration Absorber System Description}\label{sec:QCar_Model}

\subsection{System Structure}

\begin{figure}[t]
    \begin{subfigure}[h]{0.2\textwidth}
        \includegraphics[width=0.9\textwidth,center]{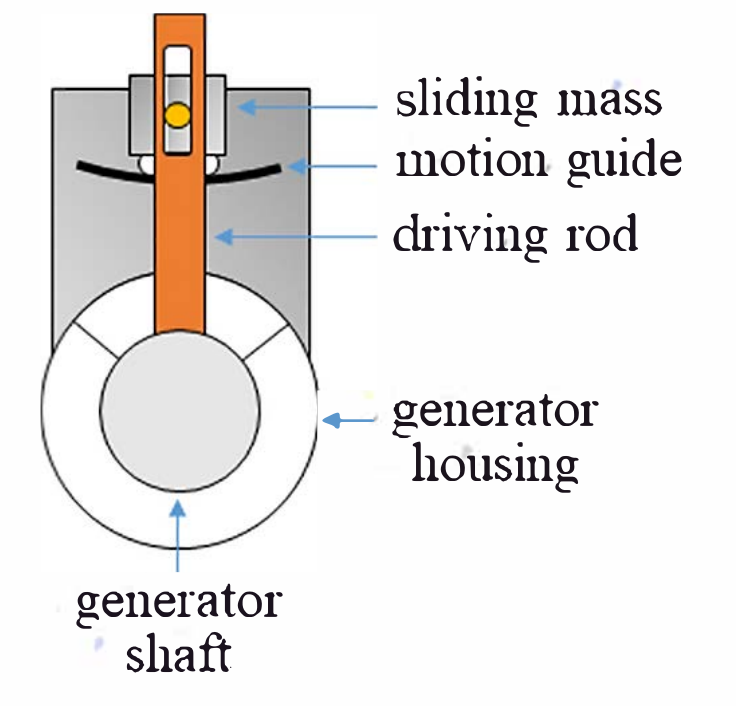}
        \label{fig1a}

    \end{subfigure}
    \begin{subfigure}[h]{0.28\textwidth}
        \includegraphics[width=1.1\textwidth,center]{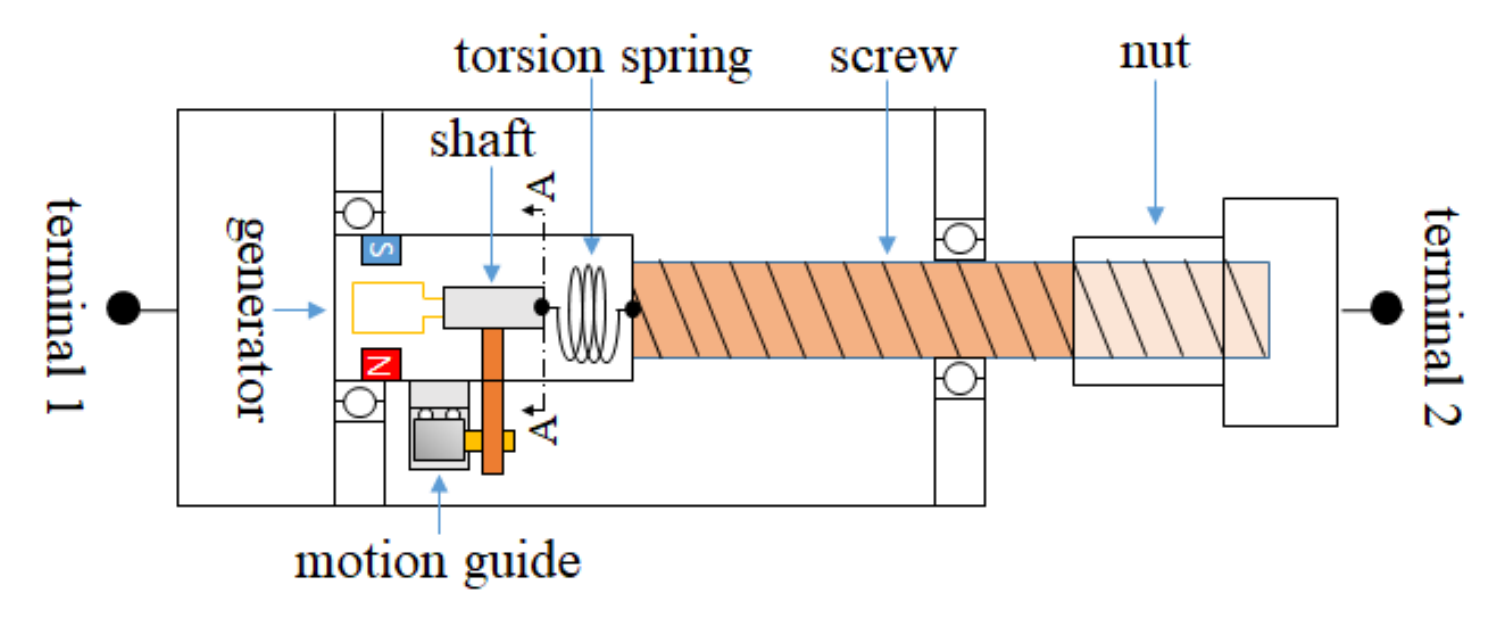}
        \label{fig1b}

    \end{subfigure}
    \caption{Schematic of the proposed electromagnetic vibration absorber with inertia nonlinearity. (left) Overview of the vibration absorber; (right) A view of the generator that passively changes its moment of inertia.}
    \label{fig1}
\end{figure}

Fig.~\ref{fig1} shows the schematic of the proposed electromagnetic vibration absorber with inertia nonlinearity. The key component of the absorber is the rotary electromagnetic generator that passively changes its moment of inertia. The passive change of moment of inertia is achieved by connecting the rotor of the electromagnetic generator with a sliding mass that is constrained to move along a motion guide fixed to the housing via a slotted rod that allows the mass to move relatively in the radial direction. Meanwhile, the angular motion of the mass and the rotor are identical. Because the radial position of the mass depends on its angular position, the moment of inertia of the rotor-mass system is a function of the angular position. Depending on the design of the motion guide, the moment of inertia can increase or decrease as the angular position changes. 
The working principle of the absorber is briefly explained as follows. When the two terminals are connected to a vehicle body and its suspension strut, respectively, the vibration of the vehicle translates the nut back and forth, subsequently rotating the screw. Because the rotor of the generator is attached to the screw via the torsion spring, it oscillates relative to the screw. As a result, some of the vibration energy is transferred to the kinetic energy of the oscillating rotor-mass system. In principle, the more kinetic energy transferred, the better the vibration suppression can be achieved. Because the kinetic energy is proportional to the moment of inertia of the absorber, one shall design a motion guide such that the moment of inertia increases as the angular position increases. Therefore, in this paper, a motion guide that enables ``enlarging inertia'' is used. 

Furthermore, the absorbed kinetic energy is converted into electricity by the electromagnetic generator to achieve energy regeneration. The rate of energy regeneration can be controlled by the duty cycle of the generator; see Fig.~\ref{fig2}. As a result, an energy regeneration circuit can be idealized as pure resistance shunt \cite{okada2002variable}. In many practical scenarios, the impedance of inductance of a generator is negligible when compared with the impedance of the energy regeneration circuit \cite{tai2018}. As such, the energy regeneration circuit can be modeled as an ideal viscous damper. By controlling the duty cycle, the damping force of the device can be controlled semi-actively.

\subsection{Equations of Motion of the Vibration Absorber}
\begin{figure}[t]
    \centering
    \includegraphics[width=0.45\textwidth]{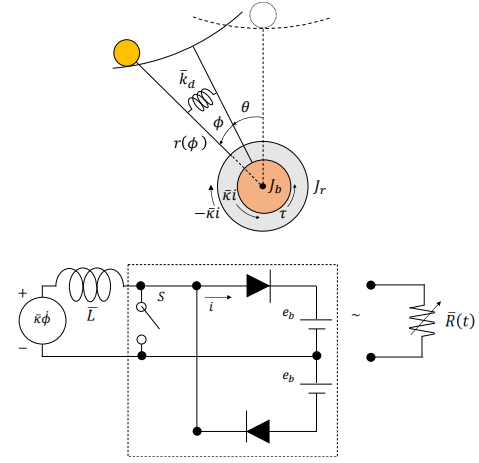}
    \caption{Mathematical model of nonlinear inertia vibration absorber}
    \label{fig2}
\end{figure}
Fig.~\ref{fig2} shows a mathematical model of the absorber. Lagrange's equation is used to derive the equations of motion of the system. Suppose the angular position of the motion guide and the angular position of the mass relative to the motion guide are $\theta$ and $\phi$, respectively. The radial position of the mass depends on the relative angular position and is written as $r=r\left(\phi\right)$. When using the polar coordinates, the kinetic energy of the system is written as,
\begin{equation}
     T=\frac{m}{2}\left[\left(r^\prime\Dot{\phi}\right)^2+r^2\left(\Dot{\phi}+\Dot{\theta}\right)^2\right]+\frac{J_r}{2}\left(\Dot{\phi}+\Dot{\theta}\right)^2+\frac{J_b}{2}\Dot{\theta}^2
     \label{eq1}
\end{equation}
where $r^\prime=dr/d\phi$ and $m$ is the mass, and $J_r$, $J_b$ are the moment of inertia of the rotor and of the ball screw (including the housing), respectively. Furthermore, the potential energy is,
\begin{equation}
     V=\frac{\Bar{k}_d}{2}\phi^2
     \label{eq2}
\end{equation}
 where $\Bar{k}_d$ is the stiffness of the torsion spring. When a pair of parallel forces of equal magnitude $F(\gg mg)$ is applied to the two terminals in opposite directions, the screw is subject to a torque $\tau=lF$, where $l$ represents the effective moment arm, which depends on the parameters of the ball screw, e.g., the screw pitch. The virtual work done by the torques is,
 \begin{equation}
     \delta W_\theta = Q_\theta\delta\theta=\tau\delta\theta
     \label{eq3}
 \end{equation}
where $Q_\theta$ is the corresponding generalized force.
The rotor and the ball screw are subject to a pair of resistance torques of equal magnitude and opposite directions due to the back electromotive force (emf). The torque on the rotor $-\Bar{\kappa}i$ is proportional to the current $i$, where $\Bar{\kappa}$ is the torque constant. The torque on the ball screw is then obtained by removing the negative sign. The energy harvesting circuit can be idealized as a variable resistance $R(t)$ that takes away the electrical energy. The virtual work done by the resistance torque is written as,
\begin{equation}
    \delta W=Q\delta \phi=-\Bar{\kappa}i\delta \phi
    \label{eq4}
\end{equation}
and the electrical virtual work done on a charge by the electrical field across the resistor is,
\begin{equation}
    \delta W_R =Q_R\delta q = -R(t)i\delta q
    \label{eq5}
\end{equation}
where $q$ is the electrical charge, and the electrical work done by the emf is,
\begin{equation}
    \delta W_{e}=Q_e\delta q=\Bar{\kappa}\Dot{\phi}\delta q
    \label{eq6}
\end{equation}
where $Q$, $Q_R$, and $Q_e$ are the generalized forces. 
Furthermore, the energy stored in the inductance is,
 \begin{equation}
     E=\frac{1}{2}Li^2
     \label{eq7}
 \end{equation}
The Lagrange equations are derived as,
\begin{equation}
    \frac{d}{dt}\frac{\partial L}{\partial \dot{g}_i}-\frac{\partial L}{\partial g_i}=Q_i
    \label{eq8}
\end{equation}
where $L=T+E-V$ is the Lagrangian, $g_1=\phi$, $g_2=\theta$, and $g_3=q$ are the generalized coordinates, and $Q_1=Q$, $Q_2=Q_\theta$, and $Q_3=Q_R+Q_e$. After substituting (\ref{eq1}) to (\ref{eq7}) into (\ref{eq8}), using the angular position of the rotor $\psi=\phi+\theta$, and idealizing the electrical circuit as pure resistance shunt, the equations of motion are derived as,

\begin{equation}\label{eq9a}
    \begin{aligned}
        & \left(J_r+J_{eff}\right)\Ddot{\psi}+mr_{eff}\Dot{\psi}^2-2mr_{eff}\Dot{\theta}\Dot{\psi}+mr^\prime r^{\prime\prime}\Dot{\theta}^2\\
        & \quad -m\left(r^\prime\right)^2\Ddot{\theta}+\Bar{k}_d\left(\psi-\theta\right)+c_e(t)\left(\Dot{\psi}-\Dot{\theta}\right)=0
     \end{aligned}
\end{equation}

\begin{equation}
     J_b\Ddot{\theta}+2m r r^{\prime}\left(\Dot{\psi}^2-\Dot{\psi}\Dot{\theta}\right)+\left(J_r+mr^2\right)\Ddot{\psi}=Fl
\end{equation}

where $J_{eff}=m\left[r^2+\left(r^\prime\right)^2\right]$ is the effective moment of inertia of the mass, $r_{eff}=\left(rr^\prime+r^\prime r^{\prime\prime}\right)$ is the effective radius, $r^{\prime\prime}=d^2r/d\phi^2$, and $c_e(t)=\Bar{\kappa}^2/R(t)$ is the semi-actively controlled electrical damping coefficient. For the rest of the paper, the radial position $r(\phi)=r(0)+\lambda r_1(\phi)=r_0+\lambda r_1(\phi)$ is considered, where $\lambda$ is a length constant.

The superior performance of this new design will be demonstrated in the numerical simulations in Section VI.A.
\section{System Formulation for Model Predictive Controller Implementation}\label{sec:Formulation}

\subsection{NMPC-based ERVA Model Formulation}

\begin{figure}[t]
    \centering
    \includegraphics[width=0.25\textwidth]{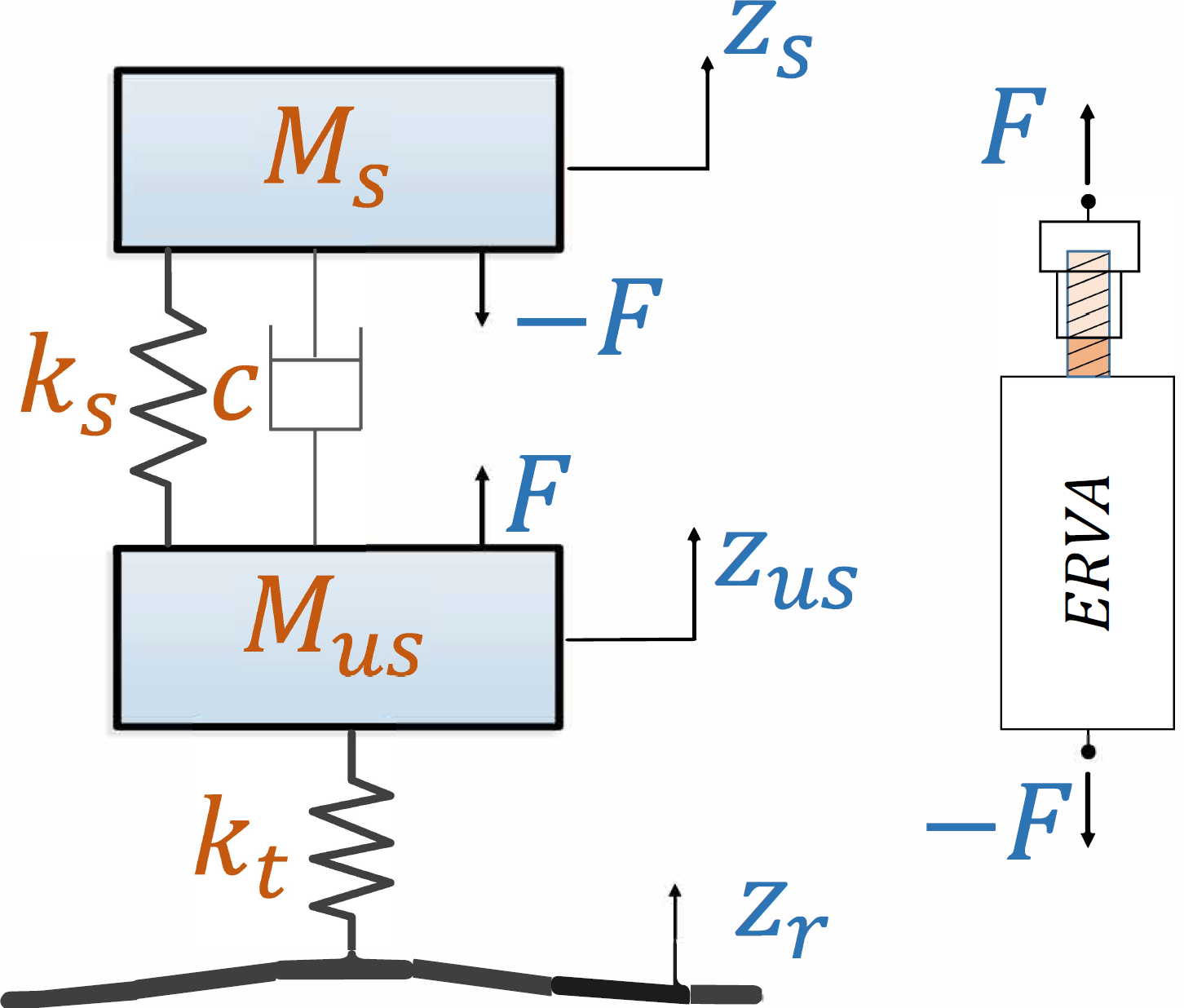}
    \caption{Quarter car model}
    \label{fig:QCar} 
\end{figure}

In this section, a nonlinear Model Predictive Controller (NMPC) is implemented for ERVA control. The system dynamics is based on the the quarter-car model as shown in Fig.~\ref{fig:QCar}. Here $M_s$ and $M_{us}$ are sprung mass and unsprung mass, respectively; $z_s$ and $z_{us}$ are the corresponding displacements, $z_r$ is the road profile; $k_s$ and $k_t$ are, respectively, spring stiffness  and tire stiffness, and $c$ is the damper coefficient. The ERVA is mounted between the sprung mass and unsprung mass, reacting by the applied force $F$. By the analysis of Newton's second law on this model, the equations of motion (EOM) of this quarter car, together with ERVA are derived in the differential equation form as,
\begin{equation}\label{equ:QCar}
\begin{aligned}
    & M_s \ddot{z}_s+m_b(\ddot{z}_s-\ddot{z}_{us})+c(\dot{z}_s-\dot{z}_{us})+k_s(z_s-z_{us})\\
    & \quad +h(x) \dot{z}_d[\dot{z}_d-(\dot{z}_s-\dot{z}_{us})]+g(x)\ddot{z}_d=0\\
    & M_{us} \ddot{z}_{us}+m_b(\ddot{z}_{us}-\ddot{z}_s)+c(\dot{z}_{us}-\dot{z}_s)+k_s(z_{us}-z_s)\\
    & \quad -h(x) \dot{z}_d[\dot{z}_d-(\dot{z}_s-\dot{z}_{us})]-g(x)\ddot{z}_d+k_t z_{us}=k_t z_r\\
    & A(x)\ddot{z}_d+B(x)\dot{z}_d^2+C(x)\dot{z}_d(\dot{z}_s-\dot{z}_{us})+D(x)(\dot{z}_s-\dot{z}_{us})^2\\
    & \quad +E(x)(\ddot{z}_s-\ddot{z}_{us})+V(z,t)+k_d[z_d-(z_s-z_{us})]=0\\
\end{aligned}
\end{equation}
where
\begin{equation}
\begin{aligned}
    & h(z)=2m_d(\epsilon r_1'+\epsilon^2 r_1 r_1')/l\\
    & g(z)=m_r+m_d(1+\epsilon r_1)^2\\
    & A(z)=m_r+m_d(1+\epsilon r_1)^2+m_d\epsilon^2(r_1')^2\\
    & B(z)=m_d[\epsilon r_1'+\epsilon^2(r_1 r_1'+r_1'r_1'')]/l\\
    & C(z)=-2B(x)\\
    & D(z)=m_d\epsilon^2r_1'r_1''/l\\
    & E(z)=-m_d\epsilon^2(r_1')^2\\
    & V(z,t)=c_e(t)[\dot{z}_d-(\dot{z}_s-\dot{z}_{us})]\\
    & r_1=\phi^2=\frac{1}{l^2}[z_d-(z_s-z_{us})]^2\\
    & r_1'=2\phi=\frac{2}{l}[z_d-(z_s-z_{us})], \quad  r_1''=2\\
\end{aligned}
\end{equation}

\begin{equation}
\begin{aligned}
\begin{array}{ccc}
     \epsilon=\frac{\lambda}{r_0}&z_s-z_{us}=l\theta& z_d=l\psi \\
     m_r=\frac{J_r}{l^2}&m_b=\frac{J_b}{l^2}&m_d=\frac{m r_0^2}{l^2}\\
     k_d=\frac{\Bar{k}_d}{l^2}&\kappa=\frac{\Bar{\kappa}}{l}\nonumber
\end{array}
\end{aligned}
\label{eq10}
\end{equation}

We define the system states as $x_1=z_s$, $x_2=\dot{z}_s$, $x_3=z_{us}$, $x_4=\dot{z}_{us}$, $x_5=z_d$ and $x_6=\dot{z}_d$; the control variable as $u=c_e(t)$; the external disturbance as $w=z_r$, then the EOMs in (\ref{equ:QCar}) can be transformed into a compact matrix form as (\ref{equ:CompM}).

\begin{equation}\label{equ:CompM}
\begin{aligned}
&\left[ \begin{array}{c c c c c c}
 1 & 0 & 0 & 0 & 0 & 0\\
 0 & M_s+m_b & 0 & -m_b & 0 & g(x)\\
 0 & 0 & 1 & 0 & 0 & 0\\
 0 & -m_b & 0 & M_{us}+m_b & 0 & -g(x)\\
 0 & 0 & 0 & 0 & 1 & 0\\
 0 & E(x) & 0 & -E(x) & 0 & A(x)\\
\end{array}
\right]
\left [ \begin{array}{c}
\dot{x}_1\\ \dot{x}_2\\ \dot{x}_3\\ \dot{x}_4\\
\dot{x}_5\\ \dot{x}_6
\end{array}
\right]=\\
&\left [ \begin{array}{ll}
  x_2\\
  -h(x)x_6(x_6-x_{2\backslash 4})-cx_{2\backslash 4}-k_sx_{1\backslash3}\\
  x_4\\
  h(x)x_6(x_6-x_{2\backslash 4})+cx_{2\backslash 4}+k_sx_{1\backslash3}-k_t x_3+k_tw\\
  x_6\\
  \left[\begin{array}{ll}
    &-B(x)x_6^2-C(x)x_6x_{2\backslash 4}-D(x)x_{2\backslash 4}^2\\
    & \quad -u(x_6-x_{2\backslash 4})-k_d(x_5-x_{1\backslash3})
 \end{array} \right]
\end{array}
\right]
\end{aligned}
\end{equation}
where $x_{2\backslash 4}=x_2-x_4$ and $x_{1\backslash3}=x_1-x_3$.

This compact matrix form $\Phi(x)\dot{x}=f(x,u,w)$, which is an implicit ordinary differential equation system, can be further re-written as $\dot{x}=\Phi(x)^{-1}f(x,u,w)$. The state-dependent coefficient matrix $\Phi(x)$, under physical system assumption, is positive definite, so the inverse of $\Phi(x)$ is applied.

The control goal for this implementation is twofold: $1)$ good ride comfort by {\it{minimizing}} the chassis acceleration $\dot{x}_2$ (which is $\ddot{z}_s$); and $2)$ efficient energy harvesting by {\it maximizing} the regenerated power, $P=u[x_6-x_{2\backslash 4}]^2$, by controlling the damping $u$ in real time. Based on the twofold target, an economic running cost function can be defined as,
\begin{equation}\label{equ:RunningCost}
    \mathcal{I}(t)=\alpha_1 x_a^2(t)-\alpha_2 u(t)\big[x_6-x_{2\backslash 4}(t)\big]^2
\end{equation}
where $x_a=\dot{x}_2$, $\alpha_1$ and $\alpha_2$ are the weights for normalizing and penalizing the twofold objective terms. So the problem considered can be formulated as the following optimal control problem,
\begin{equation}\label{equ:costJ}
\begin{aligned}
    \quad \quad \quad & min \quad J = \int_{t_0}^{t_1} \mathcal{I}(\tau)d\tau\\
    & s.t. \quad x(t_0) = x_0; \; u(t)\in \mathbb{U}\\
    & \dot{x}(t)=\Phi\big(x(t)\big)^{-1}f\big(x(t),u(t),w(t)\big)
\end{aligned}
\end{equation}
Where $t_0,\; t_1$ are the initial and terminal time points, respectively; $x_0$ is the initial condition; and $\mathbb{U}$ is the constraint set for the control variable.

In this paper, we use MPC to solve the above optimal control problem. Specifically, we discretize the continuous EOM (\ref{equ:CompM})  as $x_{k+1}=F_d(x_k,u_k,w_k)$ with a sampling time of $T_s$. The MPC problem can be formulated as,
\begin{equation}\label{equ:costJ1}
\begin{aligned}
     \quad \quad & min \quad  J = \sum_{k=0}^{N-1} \mathcal{I}_d(k)\\
    & s.t. \quad x(0) = x_0; \; u(k)\in \mathbb{U}_d\\
    & x(k+1)=\big[\Phi\big(x(k)\big)^{-1}f\big(x(k),u(k),w(k)\big)\big]_d
\end{aligned}
\end{equation}
where the discrete stage cost is $\mathcal{I}_d=\frac{\alpha_1}{T_s^2}\big[ x_a(k)-x_a(k-1) \big]^2-\alpha_2 u(k)\big[x_6(k)-x_{2\backslash 4}(k)\big]^2$; $N$ is prediction horizon; $k$ is the current time step; the EOM and control constraints are in the corresponding discrete verison. Note that with the advent of technologies that use existing onboard sensors for road profile estimation \cite{profile1,profile2}, information about the disturbance $w$ can be incorporated in the MPC framework as a preview. The above nonlinear MPC problem can be reduced to a nonlinear programming problem that can be numerically solved.

\subsection{Linear Benchmark Model}
A linear benchmark model can be obtained by considering a traditional linear rotary system, e.g., \cite{li2013}, wherein the ball screw is rigidly connected to the rotary generator. The EOMs can be readily derived from (\ref{equ:CompM}) by omitting all nonlinear terms and $z_d$'s equation.  

\begin{equation}\label{lequ:CompM}
\begin{aligned}
&\left[ \begin{array}{c c c c}
 1 & 0 & 0 & 0\\
 0 & M_s+m_x & 0 & -m_x\\
 0 & 0 & 1 & 0\\
 0 & -m_x & 0 & M_{us}+m_x\\
\end{array}
\right]
\left [ \begin{array}{c}
\dot{x}_1\\ \dot{x}_2\\ \dot{x}_3\\ \dot{x}_4
\end{array}
\right]=\\
&\left [ \begin{array}{ll}
  x_2\\
  (c+u)(x_4-x_2)+k_s(x_3-x_1)\\
  x_4\\
  (c+u)(x_2-x_4)+k_s(x_1-x_3)-k_t x_3+k_t w\\
\end{array}
\right]
\end{aligned}
\end{equation}
where $m_x=m_b+m_r+m_d$. This linear model has the fixed moment of inertia, so the advantages of ERVA with adaptive nonlinear rotational inertia is unlikely to be achieved by this linear model. However, this linear model can be used as a benchmark, representing the conventional systems with fixed rotational inertia, for performance comparison with ERVA.

\section{Simulation Results}\label{sec:Sim}
In this section, the performance of vibration control and energy harvesting is demonstrated by a set of simulations through NMPC-based ERVA. First the performance of benchmark model and passive ERVA model which has fixed electrical damping coefficient (the control variable) is examined. Then further simulations are conducted on NMPC-based ERVA model, the purpose is to find out how much the performance can be improved by knowing the road profile ahead of time. Simulation parameters are presented in Tab.~\ref{tab2}.

\subsection{Baseline vs. ERVA}
\begin{figure}[t]
    \includegraphics[width=0.4\textwidth,center]{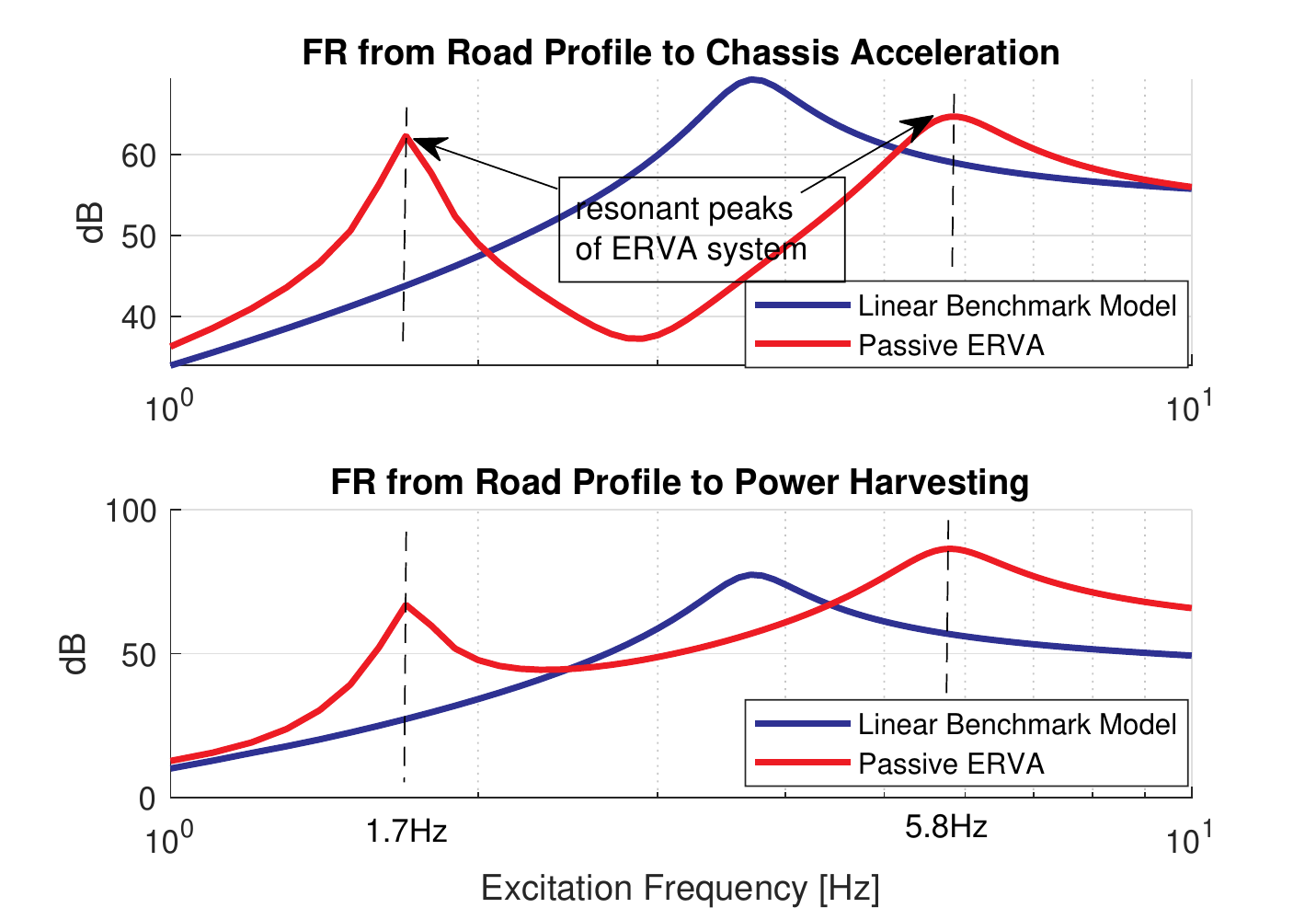}
    \caption{Approximate FR of linear benchmark and passive ERVA from road profile to; (Top) Chassis acceleration; (Bottom) Power harvesting. $1.7Hz$ and $5.8Hz$ are the two resonant peaks for passive ERVA.}
    \label{fig:fr} 
\end{figure}

In this subsection, we inspect the approximate open-loop frequency response (FR) from the road-disturbance to the chassis vertical acceleration and power harvesting system. A comparison is made between the passive ERVA with nonlinear inertia and the benchmark system described in Section III.B. The external excitation (road disturbance) is specified as a series of pure sinusoidal tones with frequency ranging from 1Hz to 10Hz. Fig.~\ref{fig:fr} shows the FR of chassis acceleration and power harvesting of the linear benchmark model and ERVA with constant control variable $(C_{e0}=10 kNs/m)$. For chassis acceleration, passive ERVA provides a remarkable attenuation benefit around the resonance frequency (located at 3.7Hz) of the benchmark model. For power harvesting, a broader bandwidth enables passive ERVA to possibly gain higher energy. The two aspects demonstrate the superior performance of the proposed ERVA design.

\begin{table}
\caption{Simulation parameters.}
\centering
\begin{tabular}{|c||c||c||c|}
\hline
\hline
$M_s$ & $M_{us}$ & $k_t$ & $k_s$\\
 \hline
$250\,\text{ kg}$ & $35\,\text{ kg}$ & $150\,\text{ kN}/\text{m}$  & $55\,\text{ kN}/\text{m}$ \\
\hline
$c$ & $k_d$ & $l$ & $c_{e0}$ \\
\hline
$70.71\, \text{ Ns}/\text{m}$ & $24.09\, \text{ kN}/\text{m}$ & $0.16\, \text{ m}/\text{rad}$ & $10\, \text{ kNs}/\text{m}$\\
\hline
$m_d$ & $m_r$ & $m_b$ & $T_s$\\
\hline
$129.78\,\text{ kg}$ & $14.42\, \text{ kg}$ & $21.31\, \text{ kg}$ & $0.02s$\\
\hline
$t_0$ & $t_1$ & $N$ & $\epsilon$\\
\hline
$0\,\text{ s}$ & $4\,\text{ s}$ & $10$ & $0.1$\\
\hline
\hline
\end{tabular}
\label{tab2}
\end{table}

\subsection{Performance of NMPC-based ERVA}
In this subsection, we directly present the results of the ERVA performance with NMPC controls, better performance of this semi-active ERVA can be easily demonstrated when compared with passive ERVA. In addition, alternative targets, i.e. minimize chassis acceleration, maximize power harvesting or trade-off according to specific requirements, are attainable with NMPC control.

For the purpose of further performance enhancement of the semi-active ERVA system. We evaluate it in two cases: 1) with no road profile preview; and 2) with road profile preview (that can be obtained using methods in \cite{profile1,profile2}. The ``true'' road profile in this simulation  is defined as a sinusoidal wave with the first natural frequency $1.7Hz$ and the wave magnitude $1mm$. In addition, a low-pass filtered unknown white-noise perturbation with Signal-to-Noise-Ratio (SNR) $7dB$ is applied. This road corresponds to a Type C road surface \cite{con33}.

\begin{figure}[!h]
    \includegraphics[width=0.4\textwidth,center]{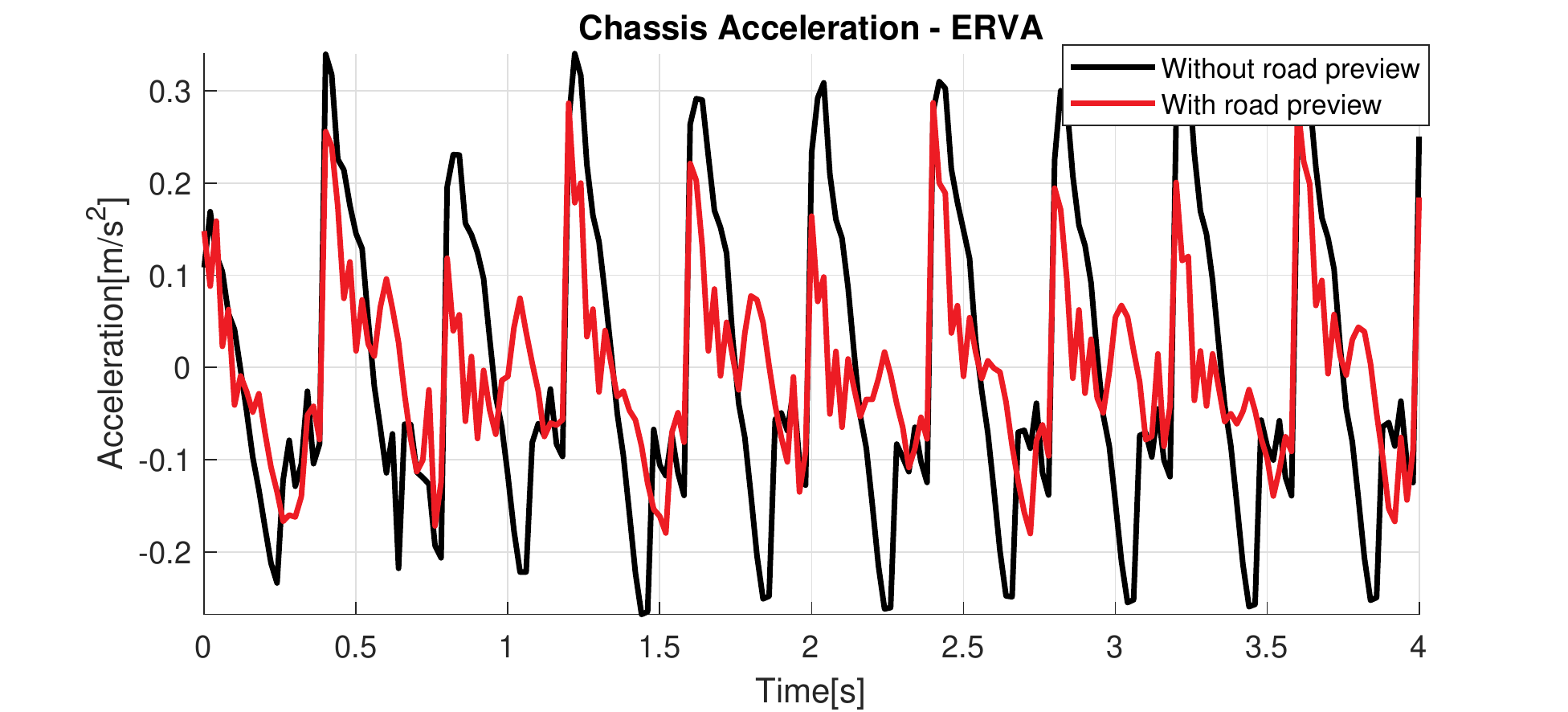}
    \caption{Minimize chassis acceleration of quarter car model.}
    \label{fig:MPC_acc} 
\end{figure}

Note that the tradeoff between energy harvesting and ride comfort can be controlled using the weights in the MPC cost function, i.e., $\alpha_1$ and $\alpha_2$ in (\ref{equ:RunningCost}).  Towards that end, we first maximize the ride comfort by minimizing the vertical acceleration  with $(\alpha_1,\alpha_2)=(1,0)$. The vertical acceleration comparison is shown in Fig.~\ref{fig:MPC_acc}, which clearly shows that MPC with preview offers better ride comfort with a 41.13\% decrease in 2-norm than MPC without preview. Note that both perform much better than the passive case which exhibits almost $1.5m/s^2$ acceleration at its first natural frequency point.

\begin{figure}[!h]
    \includegraphics[width=0.4\textwidth,center]{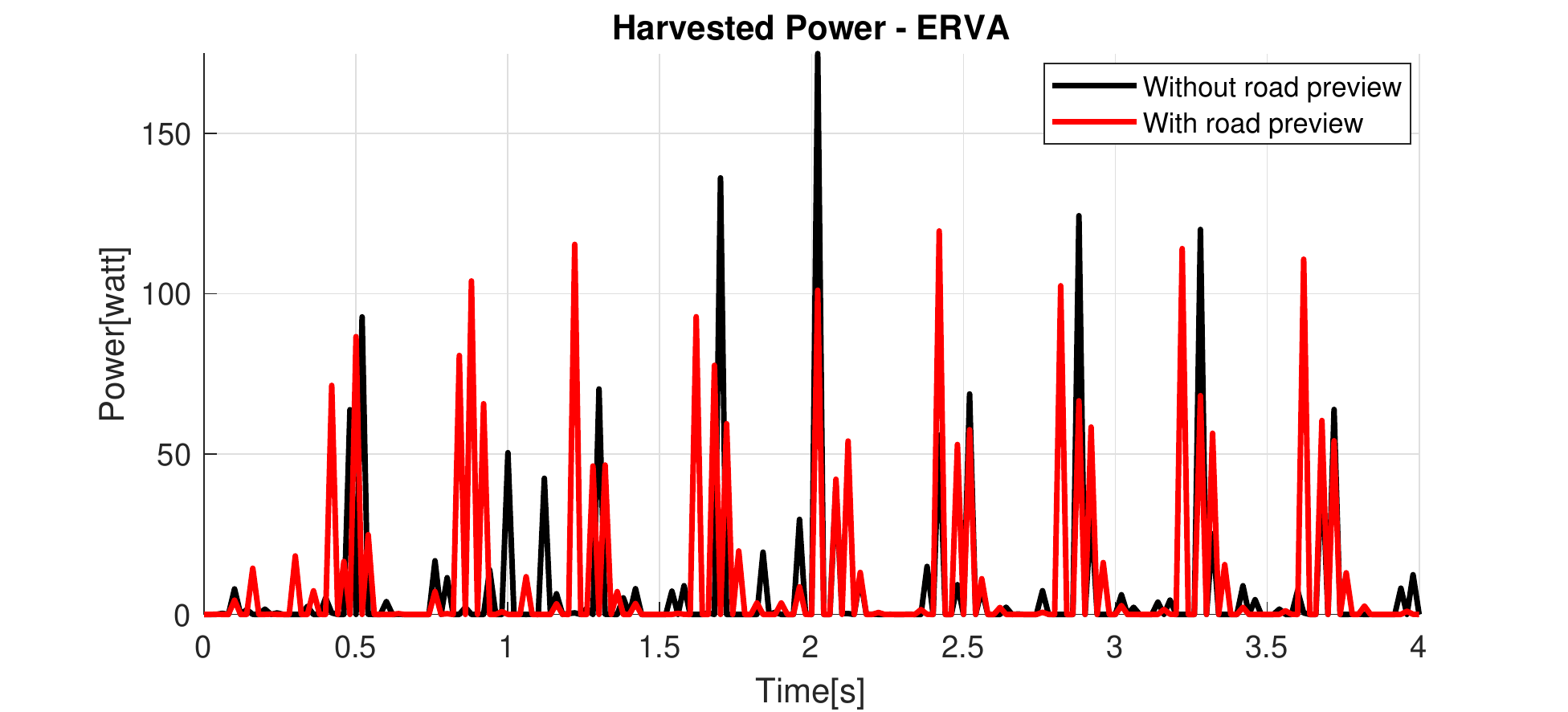}
    \caption{Maximize energy harvesting.}
    \label{fig:MPC_pow} 
\end{figure}

We next compare the energy harvesting efficiency by choosing $(\alpha_1,\alpha_2)=(0,1)$, i.e., ignoring the ride comfort term in the cost function. The performance comparison is shown in Fig.~\ref{fig:MPC_pow}. The average power regenerated by NMPC with road preview is 11.0352 watt, which is 49.13\% more compared to the no-preview case (7.3998 watt). Again, both cases for harvested power is much higher than the passive ERVA which only regenerated 2.5358 watt within the same period.

\begin{figure}[!h]
    \includegraphics[width=0.4\textwidth,center]{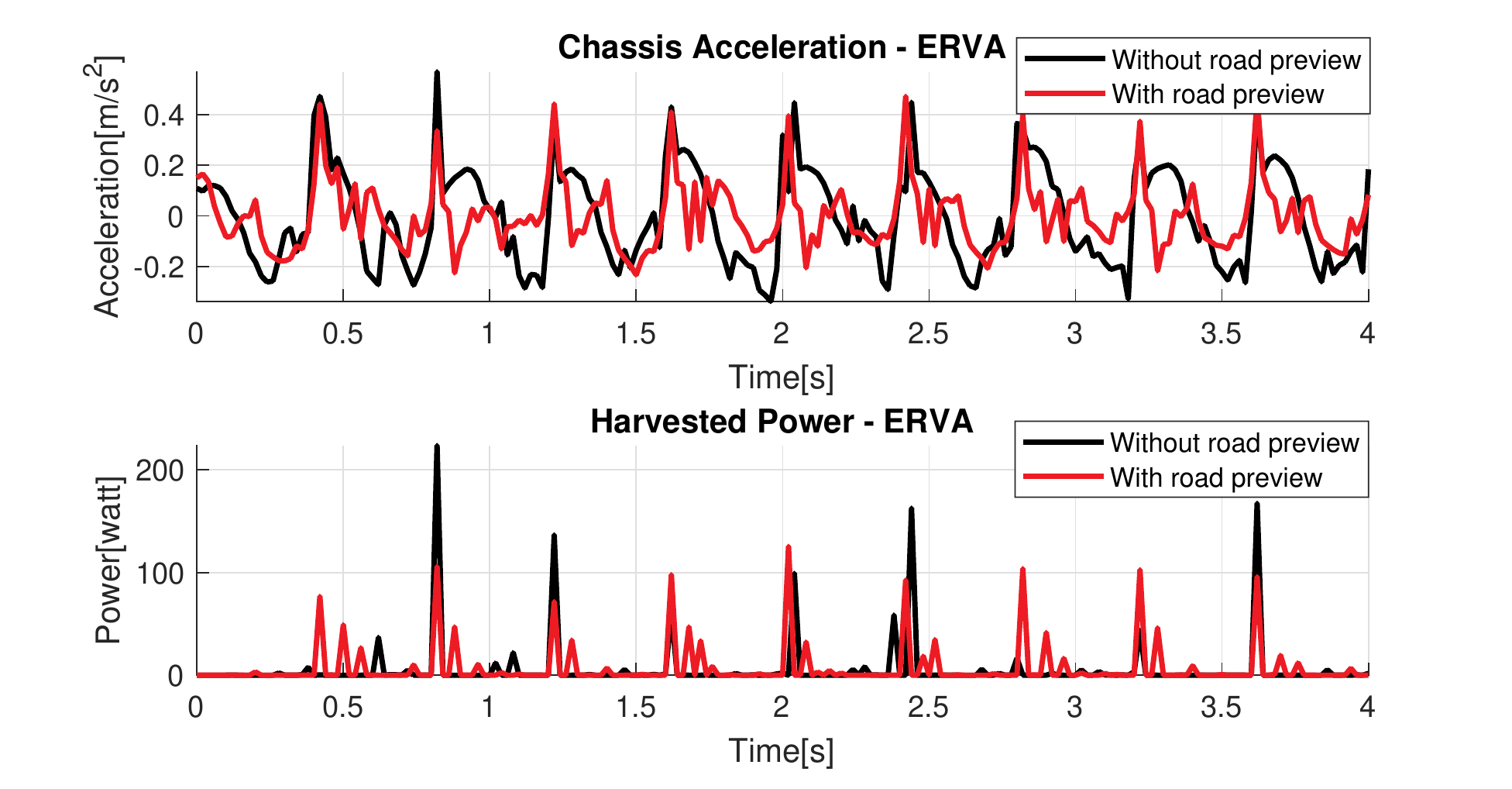}
    \caption{Twofold objective performance.}
    \label{fig:MPC_both} 
\end{figure}

Finally, both energy harvesting and ride comfort are considered in the cost function where we pick $(\alpha_1,\alpha_2)=(1,0.01)$. The comparison is shown in Fig.~\ref{fig:MPC_both}, where MPC with preview works better in terms of energy harvesting and ride comfort (26.45\% increase in power and 31.41\% decrease in vibration).

These results definitely show the benefits of incorporating road profile information as a preview, even with a moderate noise level ($7dB$ in the simulation).

\section{Conclusions}\label{sec:Conclusion}
The MPC-based electromagnetic ERVA is demonstrated effectively for vibration control and energy harvesting. This absorber with adjustable nonlinear moment of inertia can indirectly enlarge electricity converting rate by controlling the duty cycle of electromagnetic generator circuit. We demonstrate the superior performance of the ERVA as compared to a benchmark.  A nonlinear MPC is properly implemented  and we show that road preview information can be incorporated as a preview to significantly improve the control performance. Future work will focus on the stability and sensitivity analysis of NMPC  as well as prototyping the proposed ERVA.


\bibliographystyle{IEEEtran}
\bibliography{IEEEexample}

\end{document}